\RequirePackage{lineno}
\documentclass[aps,prc,twocolumn, floatfix]{revtex4}
\usepackage{graphicx}
\usepackage{dcolumn}
\usepackage{bm}
\usepackage{multirow}
\usepackage[colorlinks=true, linkcolor=blue, citecolor=blue, urlcolor=blue]{hyperref}

\begin{document}


\title{A direct probe of $\Lambda$ potential in nuclear medium}

\author{Gao-Chan Yong$^{1,2}$}

\affiliation{
$^1$Institute of Modern Physics, Chinese Academy of Sciences, Lanzhou 730000, China\\
$^2$School of Nuclear Science and Technology, University of Chinese Academy of Sciences, Beijing 100049, China
}

\begin{abstract}

Using the Li\`{e}ge intranuclear-cascade model together with the ablation model ABLA, an investigation is conducted into the effects of $\Lambda$ potential in $\Lambda$-nucleus and $\Lambda$-hypernucleus-nucleus collisions across various beam energies. The findings show that the angle and transverse-momentum distributions of scattered $\Lambda$ hyperon, the scattering cross section of the $\Lambda$ hyperon in $\Lambda$-nucleus collisions as well as the rapidity distribution of $\Lambda$ hyperon in $\Lambda$-hypernucleus-nucleus collisions are significantly influenced by the strength of the $\Lambda$ potential in these scattering reactions across various beam energies. These demonstrations, unhindered by the uncertainties of $\Lambda$ and hypernuclei productions in nuclear medium, allow for a direct investigation of the $\Lambda$ potential, especially its momentum dependence. The criticality of probing the $\Lambda$ potential is closely associated with the resolution of the "hyperon puzzle" in neutron stars.

\end{abstract}

\maketitle

%
Neutron stars emerge from the ashes of supernova explosions and were initially believed to predominantly consist of neutrons. Nonetheless, a sequence of theoretical studies has suggested that neutron stars might also harbor strange matter \cite{Weber2005}. The concept of strangeness in neutron stars could profoundly impact our comprehension of these celestial bodies \cite{nkg2001}. This aspect might influence their internal composition and cooling mechanisms, as highlighted in various studies \cite{cool92,cool99,cool3,cool4,cool5}. Additionally, the presence of strangeness could alter the dynamics of neutron star mergers \cite{mark21}, potentially leading to distinct gravitational wave signatures \cite{gw17}. The softening of the equation of state for dense matter, attributed to strangeness, could play a pivotal role in the shock-wave/neutrino-delayed-shock process, thereby affecting the explosive birth of core-collapse supernovae \cite{npa1997,soft2,soft3,soft4}.
The presence of strangeness within neutron star matter has thus ignited the curiosity of the physics community \cite{apj85,sch96,nk82,lon15,ch16,wz12,ger20,jj23}, promising to shed light on the strong nuclear force and the characteristics of matter in ultra-extreme conditions, as further explored in references \cite{wz12,jj23,sxx2023,vid18,ppnp2020,chenjh}. Consequently, the study of hyperon-nucleon interactions and the significance of strangeness in neutron stars arises as a vital domain for astrophysicists, together with particle and nuclear physicists.

The resolution of the ``hyperon puzzle'', focusing on the inclusion of strangeness within the nuclear medium to ascertain the strong nuclear force, remains a fundamental query \cite{vid18}. A variety of theoretical approaches have been adopted, such as utilizing nuclear many-body theories for calculating the properties of strange particles in dense environments \cite{ch16,ppnp2020}, applying effective field theories to elucidate the interactions between strange particles and nucleons \cite{jm14,geng2022}, using perturbative quantum chromodynamics for the study of strange quark matter \cite{ms1987,xiacj2017}. Astrophysical observations of neutron stars also serve as a means to deduce the characteristics of strange particles within \cite{longwh12,hell14}.

Despite these methodologies, interactions between strangeness and non-strangeness in nuclear matter confront considerable theoretical ambiguities and are seldom directly examined through nuclear experiments in terrestrial labs. In Ref.~\cite{yongprd2023}, it is argued that combining studies of transport models with nuclear experimental data from facilities around the world could be a better way to address the ``hyperon puzzle''. Therefore, nuclear reactions may offer a more effective method for studying hyperon-nucleon interactions within nuclear medium: Hyperon-nucleus scattering experiments deal with hyperon-nucleon interactions in medium around saturation density, whereas hypernucleus-nucleus collisions examine these interactions at higher densities.
By adjusting the beam energy, it is possible to investigate the momentum dependence of hyperon-nucleon interactions in the nuclear medium, as hyperon potentials are momentum-dependent, utilizing, for instance, an extended Brueckner-Hartree-Fock formalism \cite{mdi1998}.
In this study, the Li\`{e}ge intranuclear-cascade (INCL) model, combined with the ABLA deexcitation code, is employed to investigate the effects of the $\Lambda$ potential around and above normal nuclear density. It's demonstrated that experiments involving $\Lambda$-nucleus and hypernucleus-nucleus collisions could directly unveil the $\Lambda$-nucleon interactions in nuclear medium.

%
The Li\`{e}ge intranuclear-cascade code, also known as the INCL model \cite{incl2002,incl13,incl14}, is used to describe the collision between a projectile (such as nucleons, pions, hyperons and light ions) and a target nucleus. The INCL model incorporates classical physics principles, but also includes some quantum-mechanical features (such as Fermi motion, realistic space densities and Pauli blocking) to account for the initial conditions and dynamics of the collision. The INCL model treats an energy and isospin-dependent nucleon potential and an isospin-dependent constant hyperon potential with Woods-Saxon density distributions \cite{incl13,incl2017,incl2018}. The model treats nuclear collisions as successive relativistic binary hadron-hadron collisions, where the positions and momenta of the hadrons are tracked over time. The extended INCL model includes the production of pion mesons and strange particles \cite{incl11pion,incl18strange}. The latest version of the model, INCL++6.32, includes the formation of hyperremnants \cite{incl20strange,incl22strange}. Based on different kinds of conservation laws (such as for baryon number, charge, energy, momentum, and angular momentum), the model predicts the formation of hot hyperremnants and characterizes them in terms of atomic and mass numbers, strangeness number, excitation energy, and angular momenta.

On the other hand, the ABLA07 model \cite{abla07} is a deexcitation model that describes the subsequent de-excitation processes after the collision.
It describes the de-excitation of an excited remnant by emitting $\gamma$ rays, neutrons, light-charged particles, and intermediate-mass fragments (IMFs) or fission decays.
The ABLA07 model was rewritten in C++ from its original FORTRAN version to be compatible with the GEANT4 simulation framework. The ABLA07 model is considered to be one of the best model for describing the deexcitation of hot remnants, according to benchmarks performed by the International Atomic Energy Agency \cite{abla07}. Recently, the ABLA07 model has been extended to include the emission of hyperons and the formation of cold hypernuclei \cite{incl22strange,prl2023}. Overall, the INCL model is used to describe the collision stage of the reactions, while the updated ABLA model is used to describe the subsequent deexcitation stage. The combination of the two models allows for the study of
hyperon-related scatterings and provides insights into the properties of hyperon in nuclear matter.

%
Scattering experiments play a pivotal role in elucidating the complex interactions among particles. These experiments provide a profound understanding of particle characteristics and find wide applications in numerous scientific fields, including nuclear physics and quantum mechanics. In particular, scattering experiments are integral to deriving the optical potential, a critical element in the field of nuclear physics. The optical potential, usually connected to the potential energy of participant particles, foretells the particle movement within a quantum system. This potential can be extracted through a careful examination of the trajectories and speeds of both outgoing and incoming particles during scattering experiments. Indeed, the $\Lambda$-nucleon potential can be scrutinized through the production of hypernuclei, which helps in gauging the \emph{momentum-independent} $\Lambda$ potential in medium around saturation density \cite{incl2018}. Likewise, studying the directed flows of hyperons or hypernuclei provides insights into the $\Lambda$ potential in medium at high densities \cite{star2023}. Notwithstanding, such research poses prominent theoretical uncertainties, primarily linked to the production of $\Lambda$ and the assembly of hypernuclei as well as non-strange baryon potential since the $\Lambda$ hyperon is in fact secondary particles in heavy-ion collisions.

\begin{figure}[t]
\centering
\vspace{-0.5cm}
\includegraphics[width=0.485\textwidth]{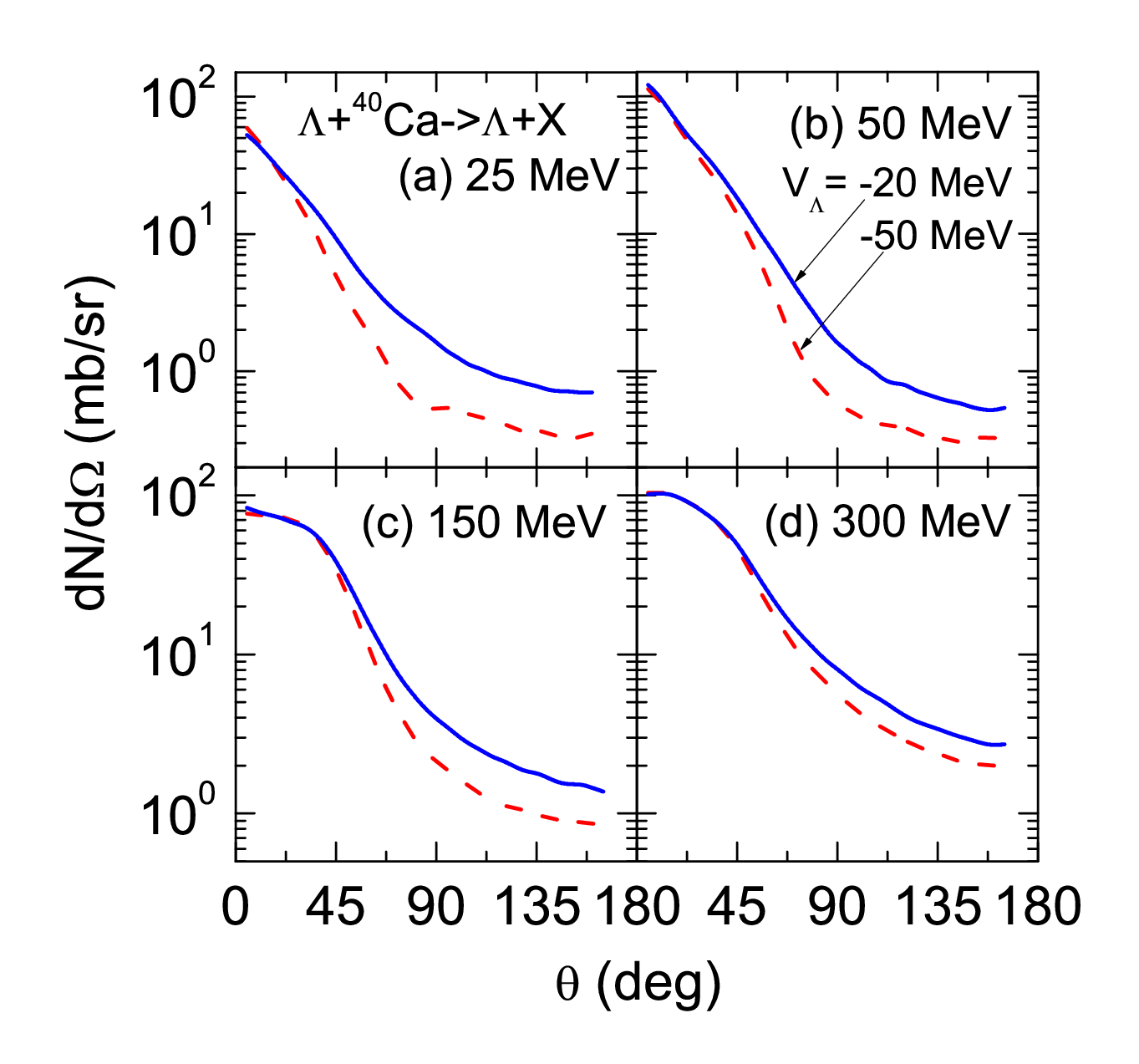}
\vspace{-0.7cm}
\caption{Angle distributions of $\Lambda$ emissions in $\Lambda$+$^{40}$Ca scatterings under varying $\Lambda$ potentials and beam energies of 25 (a), 50 (b), 150 (c), 300 (d) MeV.} \label{angle}
\end{figure}
Figure~\ref{angle} illustrates the impact of the $\Lambda$ potential on the angular distributions of $\Lambda$ emissions in $\Lambda$+$^{40}$Ca scatterings across various $\Lambda$-beam energies. The $^{40}$Ca target (together with the latter used $^{16}$O) ensures the absence of isospin effects from the $\Lambda$ potential. Given that the $\Lambda$ potential can transition from a notably negative value at very low momenta to a positive one at high momenta around saturation density, I modify the $\Lambda$ potential value from the initially estimated -30 MeV at saturation density  \cite{incl2018,incl22strange} to a span of -50 to -20 MeV at normal density for illustrative purposes. At high densities, the $\Lambda$ potential might turn positive. To maintain simplicity, this range is still employed to demonstrate the effects of the $\Lambda$ potential. Observations indicate that at narrow polar angles, the influence of the $\Lambda$ potential on $\Lambda$ emissions in $\Lambda$+$^{40}$Ca scatterings is subtle. However, at wider polar angles ($\theta$ $>$ 60), $\Lambda$ emissions exhibit pronounced effects of the $\Lambda$ potential. The less attractive the $\Lambda$ potential for $\Lambda$ hyperons (the larger the $\Lambda$ potential), the greater the number of $\Lambda$ emissions observed at these wider angles. Consequently, more $\Lambda$ emissions are observed with a $\Lambda$ potential of -20 MeV compared to -50 MeV. As the beam energy increases, the significance of the $\Lambda$ potential's effects diminishes progressively. Given that the $\Lambda$ potential may also be momentum-dependent, the beam-energy-dependent analysis presented here could serve as a means to investigate the momentum dependence of the $\Lambda$ potential in nuclear medium.

\begin{figure}[t]
\centering
\vspace{-0.25cm}
\includegraphics[width=0.5\textwidth]{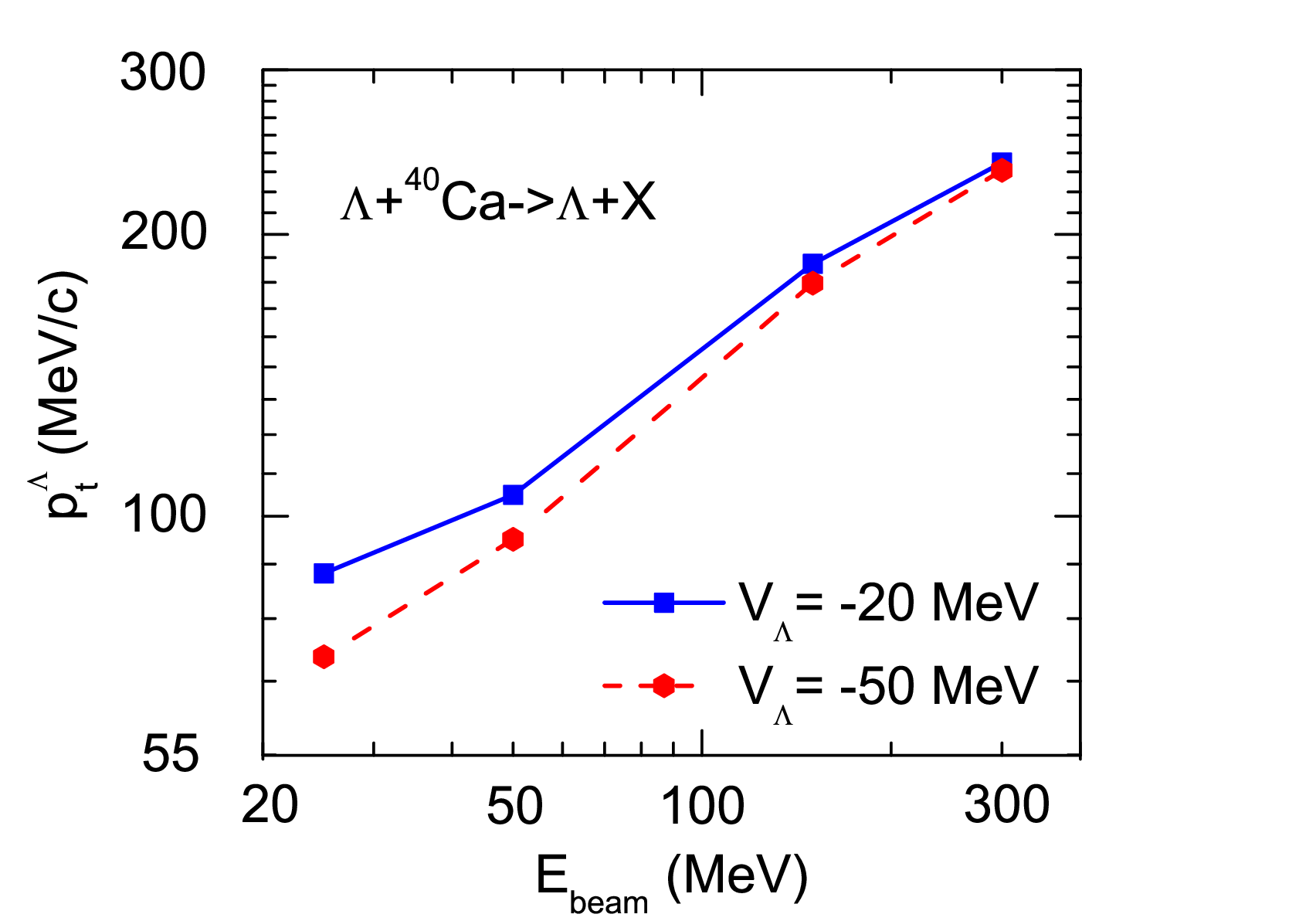}
\caption{Impact of the $\Lambda$ potential on the transverse momentum of $\Lambda$ in $\Lambda$+$^{40}$Ca scatterings at varying beam energies.} \label{pt}
\end{figure}
Considering the pronounced impact of the $\Lambda$ potential on $\Lambda$ emissions at large polar angles, it becomes significant to explore its influence on $\Lambda$ transverse momentum in $\Lambda$+$^{40}$Ca scatterings. Figure~\ref{pt} illustrates the behavior of $\Lambda$ transverse momentum under various beam energies. As anticipated, there is an observable increase in $\Lambda$ transverse momentum corresponding to the rise in beam energies. Concurrently, the influence of the $\Lambda$ potential on this transverse momentum diminishes as the beam energies escalate. This trend is logical, given that at higher energies, the $\Lambda$ potential exerts a less noticeable effect on $\Lambda$+$^{40}$Ca scatterings. Additionally, Figure~\ref{pt} highlights that an increase in the $\Lambda$ potential (rendering it less attractive) leads to higher $\Lambda$ transverse momenta. Hence, the $\Lambda$ transverse-momentum spectra could serve as a viable tool for assessing the strength of the $\Lambda$ potential in $\Lambda$+$^{40}$Ca scatterings.

\begin{figure}[t!]
\centering
\vspace{-0.1cm}
\includegraphics[width=0.4\textwidth]{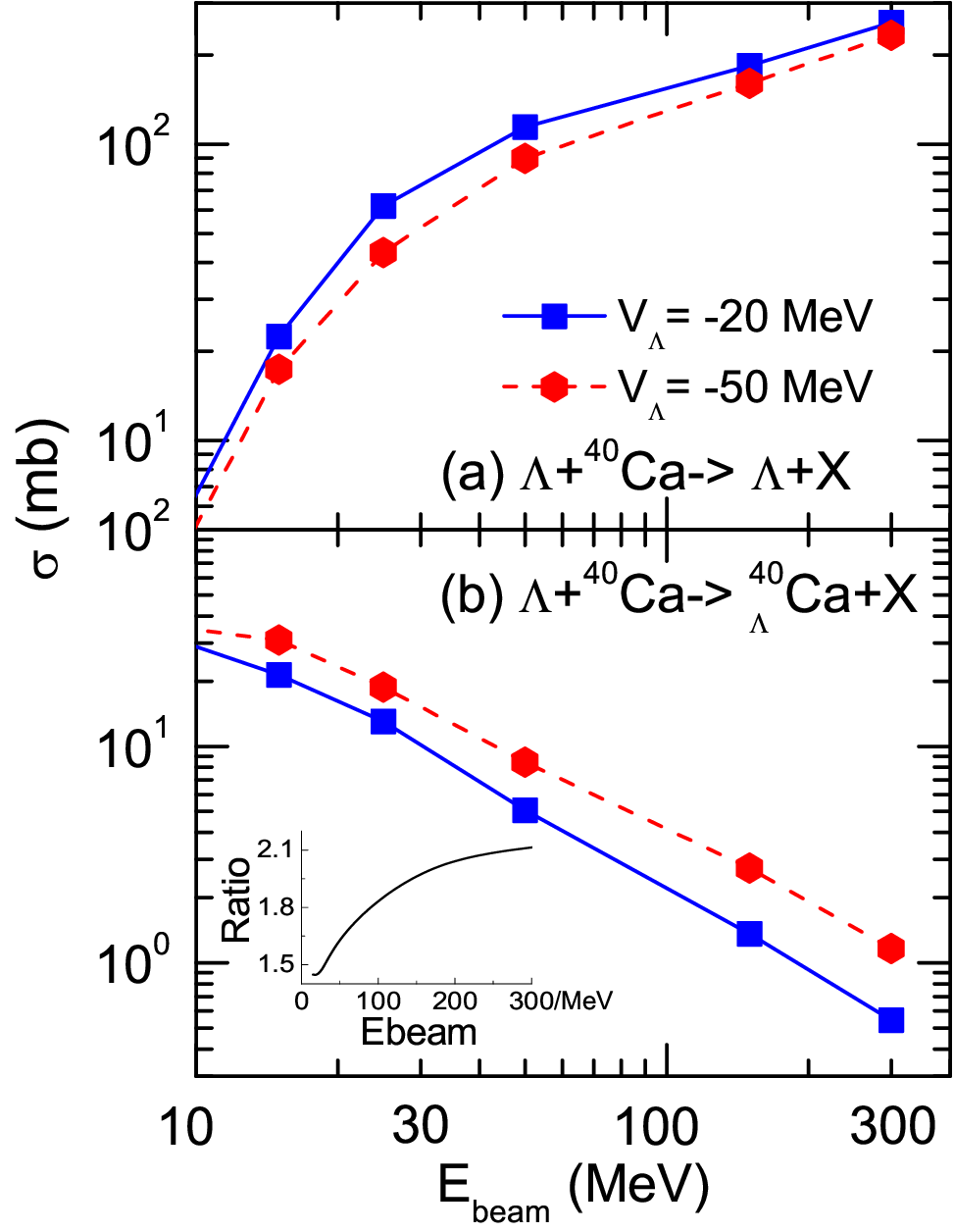}
\caption{Total cross sections for $\Lambda$ (a) and hypernucleus $^{40}_{\Lambda}$Ca (b) productions in $\Lambda$+$^{40}$Ca scatterings across various beam energies and $\Lambda$ potentials. The inset shows the ratio (Ratio = $\sigma^{^{40}_{\Lambda}Ca}_{-50}/\sigma^{^{40}_{\Lambda}Ca}_{-20}$) of $^{40}_{\Lambda}$Ca production cross sections with the two $\Lambda$ potentials as a function of beam energy.} \label{xsection}
\end{figure}

To investigate the momentum dependence of the $\Lambda$ potential around saturation density, examining the beam-energy dependent $\Lambda$ scattering cross section in the $\Lambda$+$^{40}$Ca reaction is one approach. Panel (a) of Figure~\ref{xsection} presents the $\Lambda$ scattering cross section across various beam energies and $\Lambda$ potentials. It's observed that the total cross sections for $\Lambda$ scatterings escalate as the beam energy increases. This increase is attributed to the diminished capturing capability of the $\Lambda$ potential for higher energy $\Lambda$ particles in the $^{40}$Ca target. The influence of the $\Lambda$ potential on the total cross sections of $\Lambda$ scatterings becomes less significant as the beam energy rises. A larger (less attractive) $\Lambda$ potential leads to higher total $\Lambda$ production cross sections, aligning with the findings depicted in Figure~\ref{angle}.

Panel (b) of Figure~\ref{xsection} illustrates the total cross sections of hypernucleus $^{40}_{\Lambda}$Ca production as a function of beam energy. Given that the total cross sections for $\Lambda$ scatterings rise with increasing beam energy, an inverse trend in the total cross sections of hypernucleus $^{40}_{\Lambda}$Ca production with increasing beam energy is observed, which is logical. Intriguingly, the impact of the $\Lambda$ potential on the total cross sections of hypernucleus $^{40}_{\Lambda}$Ca production intensifies with increasing beam energy, which can be clearly seen from the inset of Panel (b). This is because a higher energy $\Lambda$ particle requires a greater number of scatterings -- resulting in a longer duration within the target and a heightened influence from the $\Lambda$ potential -- to decelerate and eventually be captured by the $\Lambda$ potential in the target $^{40}$Ca. Hence, the total cross sections of hypernucleus $^{40}_{\Lambda}$Ca production in $\Lambda$+$^{40}$Ca scatterings serve as a valuable means to explore the momentum-dependent $\Lambda$ potential around saturation density.

The investigation into the $\Lambda$ hyperon potential in dense nuclear environments is critically important for understanding neutron stars, especially when it comes to solving the ``hyperon puzzle''. This puzzle arises from the contradiction between theoretical models, which predict that hyperons would soften the equation of state (EoS), leading to the formation of lower mass neutron stars, and the recent observations of neutron stars with significantly higher masses. Research focused on the role of the $\Lambda$ hyperon seeks to align the EoS with the existence of these more massive neutron stars. Therefore, exploring the high-density $\Lambda$ potential sheds light on the behavior of matter under extreme conditions and could be key to resolving the hyperon puzzle.

\begin{figure}[t]
\centering
\vspace{-0.25cm}
\includegraphics[width=0.5\textwidth]{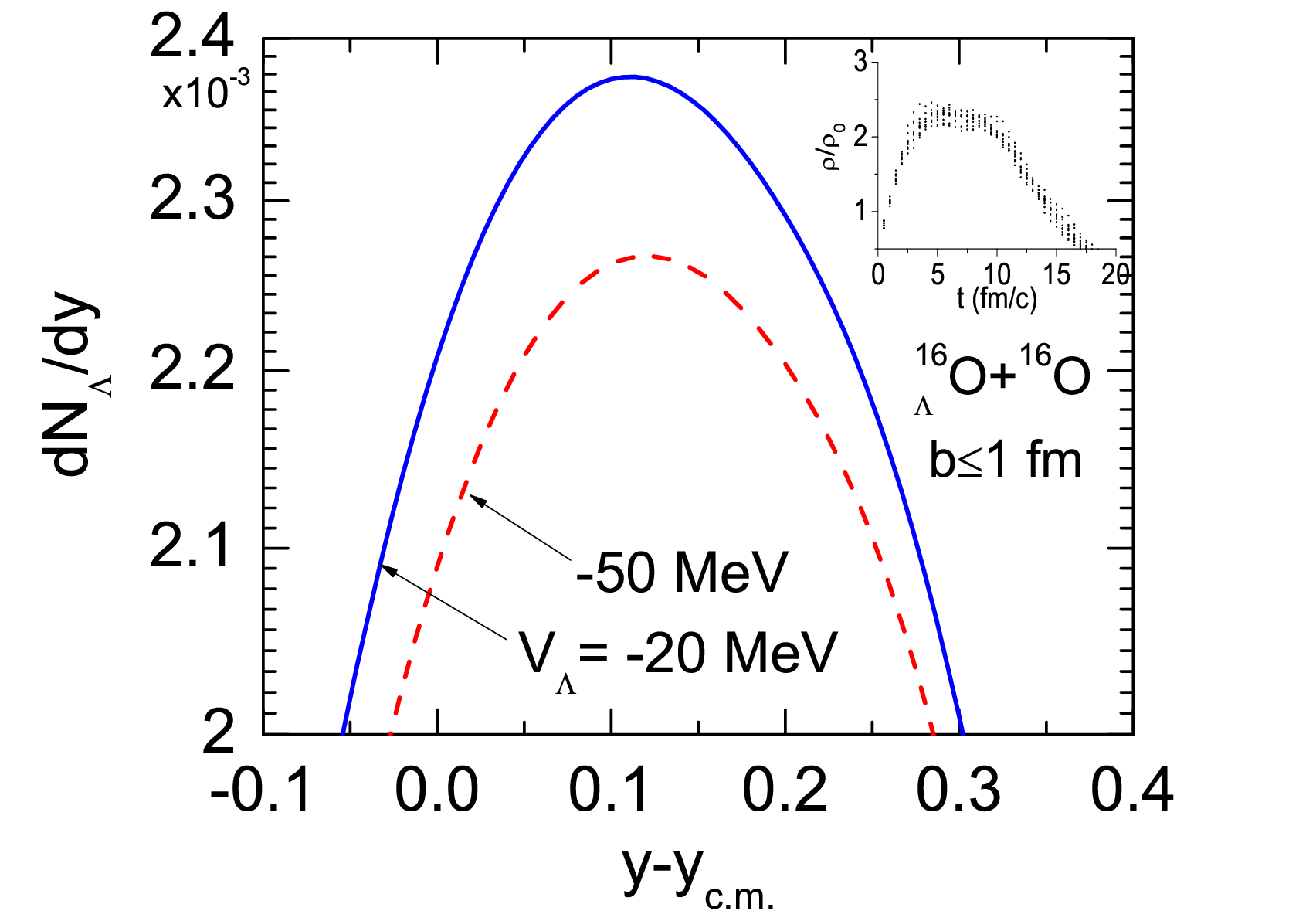}
\caption{Rapidity distributions for $\Lambda$ production in central $^{16}_{\Lambda}$O+$^{16}$O collisions at 300 MeV/nucleon under various $\Lambda$ potentials. The inset illustrates the central maximum compression density as a function of reaction time.} \label{highden}
\end{figure}
To replicate high-density nuclear matter in terrestrial labs, heavy-ion collisions are frequently employed. Figure~\ref{highden} illustrates the central $^{16}_{\Lambda}$O+$^{16}$O collisions at 300 MeV/nucleon under varying $\Lambda$ potentials. The inset reveals that dense matter, approximately twice the saturation density, is achieved in these central collisions at 300 MeV/nucleon. The figure shows how the $\Lambda$ hyperon's rapidity distributions in these collisions are significantly influenced by the $\Lambda$ potential. A larger (less attractive) $\Lambda$ potential results in increased $\Lambda$ emissions at mid-rapidities. Given the $\Lambda$ hyperon's presence exclusively in the projectile, it predominantly appears in the positive-rapidity region. Should the $\Lambda$ potential be positive in high-density nuclear environments \cite{jinno2023}, the impact on the rapidity distributions of $\Lambda$ production would be markedly greater.

The reactions involving $\Lambda$ can be conducted through secondary-beam experiments in terrestrial laboratories, such as those performed at JLab/CLAS \cite{clas2019} and BEPCII/BESIII \cite{bas31,bas32}. Reactions involving $\Lambda$-hypernuclei could potentially be executed at GSI/FAIR, using a dedicated HYDRA (HYpernuclei Decay at R3B Apparatus) time-projection chamber prototype \cite{gsi23}. Given that current studies bypass the generations of $\Lambda$-hyperons or $\Lambda$-hypernuclei in the medium, theoretical uncertainties are substantially minimized.

%
To summarize, utilizing the Li\`{e}ge intranuclear-cascade model in tandem with the ABLA deexcitation model, this research delves into the dynamics of $\Lambda$-nucleus and hypernucleus-nucleus interactions. It aims to uncover how the $\Lambda$ potential influences $\Lambda$ particles in nuclear medium. Findings indicate that in the $\Lambda$+$^{40}$Ca collisions, $\Lambda$ emissions are markedly influenced by the $\Lambda$ potential, particularly at wide polar angles. Moreover, the impact of the $\Lambda$ potential on the $\Lambda$ particle's transverse momentum is distinctly observable. While the influence of the $\Lambda$ potential on total cross sections of $\Lambda$ scatterings diminishes as the beam energy escalates, its effect on the total cross sections for the production of the hypernucleus $^{40}_{\Lambda}$Ca intensifies with increasing beam energy. The rapidity distributions of $\Lambda$ production in central $^{16}_{\Lambda}$O+$^{16}$O collisions at 300 MeV/nucleon significantly reflect the $\Lambda$ potential's effects, offering a potential method to probe the $\Lambda$ potential at high densities. This investigation aims at directly probing the $\Lambda$ potential in the nuclear medium and addressing the ``hyperon puzzle'' in neutron stars.

%
The author thanks J. L. Rodr\'{\i}guez-S\'{a}nchez, J. Cugnon, J.-C. David and D. Mancusi for helpful communications. This work is supported by the National Natural Science Foundation of China under Grant Nos. 12275322, 12335008.

\end{document}